\documentclass[sigconf]{acmart}

\AtBeginDocument{%
  
}

\copyrightyear{2025}
\acmYear{2025}
\setcopyright{cc}
\setcctype{by-nc-nd}
\acmConference[ICMI '25]{Proceedings of the 27th International Conference on Multimodal Interaction}{October 13--17, 2025}{Canberra, ACT, Australia}
\acmBooktitle{Proceedings of the 27th International Conference on Multimodal Interaction (ICMI '25), October 13--17, 2025, Canberra, ACT, Australia}

\usepackage{xcolor}
\usepackage{amsmath}

\usepackage{tabularx}
\usepackage{graphicx}
\usepackage{subcaption}
\usepackage{multirow}
\usepackage[inline]{enumitem} 
\usepackage{xspace}
\setlist[itemize]{leftmargin=*, nosep}

\newcommand{\jpei}[1]{\textcolor{black}{#1}}

\newcommand{\mypara}[1]{\smallskip\noindent\textbf{#1.}\xspace}

\usepackage{acronym}
\acrodef{AI}{artificial intelligence}
\acrodef{LLM}{large language model}

\begin{document}

\title{\jpei{Talking-to-Build: How LLM-Assisted Interface Shapes Player Performance and Experience in Minecraft}}


\author{Xin Sun}
\affiliation{\institution{University of Amsterdam, NL}
  \country{National Institute of Informatics (NII)}, JP}
\authornote{Both authors contributed equally to this research.}

\author{Lei Wang}
\affiliation{
  \institution{Vrije Universiteit Amsterdam}
  \country{Universiteit Utrecht, \\NL}}
\authornotemark[1]

\author{Yue Li}
\affiliation{\institution{Vrije Universiteit Amsterdam}
  \country{NL}}

\author{Jie Li}
\affiliation{%
  \institution{Human-AI Symbiosis Alliance}
  \country{CH}}

\author{Massimo Poesio}
\authornotemark[3]
\authornotemark[6]
\author{Julian Frommel}
\authornotemark[3]
\affiliation{%
  \institution{\ddag Universiteit Utrecht, NL}
  \country{$\parallel$ Queen Mary University, UK}}

\author{Koen Hinriks}
\author{Jiahuan Pei}
\authornote{Corresponding author: jpei2@vu.nl}
\affiliation{%
  \institution{Vrije Universiteit Amsterdam}
  \country{NL}
}



\begin{abstract}
\jpei{
With large language models (LLMs) on the rise, in-game interactions are shifting from rigid commands to natural conversations.
However, the impacts of LLMs on player performance and game experience remain underexplored.
This work explores LLM's role as a co-builder during gameplay, examining its impact on task performance, usability, and player experience.
Using Minecraft as a sandbox, we present an LLM-assisted interface that engages players through natural language, aiming to facilitate creativity and simplify complex gaming commands.
We conducted a mixed-methods study with 30 participants, comparing LLM-assisted and command-based interfaces across simple and complex game tasks. 
Quantitative and qualitative analyses reveal that the LLM-assisted interface significantly improves player performance, engagement, and overall game experience.
Additionally, task complexity has a notable effect on player performance and experience across both interfaces.
Our findings highlight the potential of LLM-assisted interfaces to revolutionize virtual experiences, emphasizing the importance of balancing intuitiveness with predictability, transparency, and user agency in AI-driven, multimodal gaming environments.
}
\end{abstract}


\vspace{-4mm}


\vspace{-2mm}


\maketitle


    








\section{Introduction}

\begin{figure}[ht!]
    \centering
\includegraphics[width=.99\linewidth]{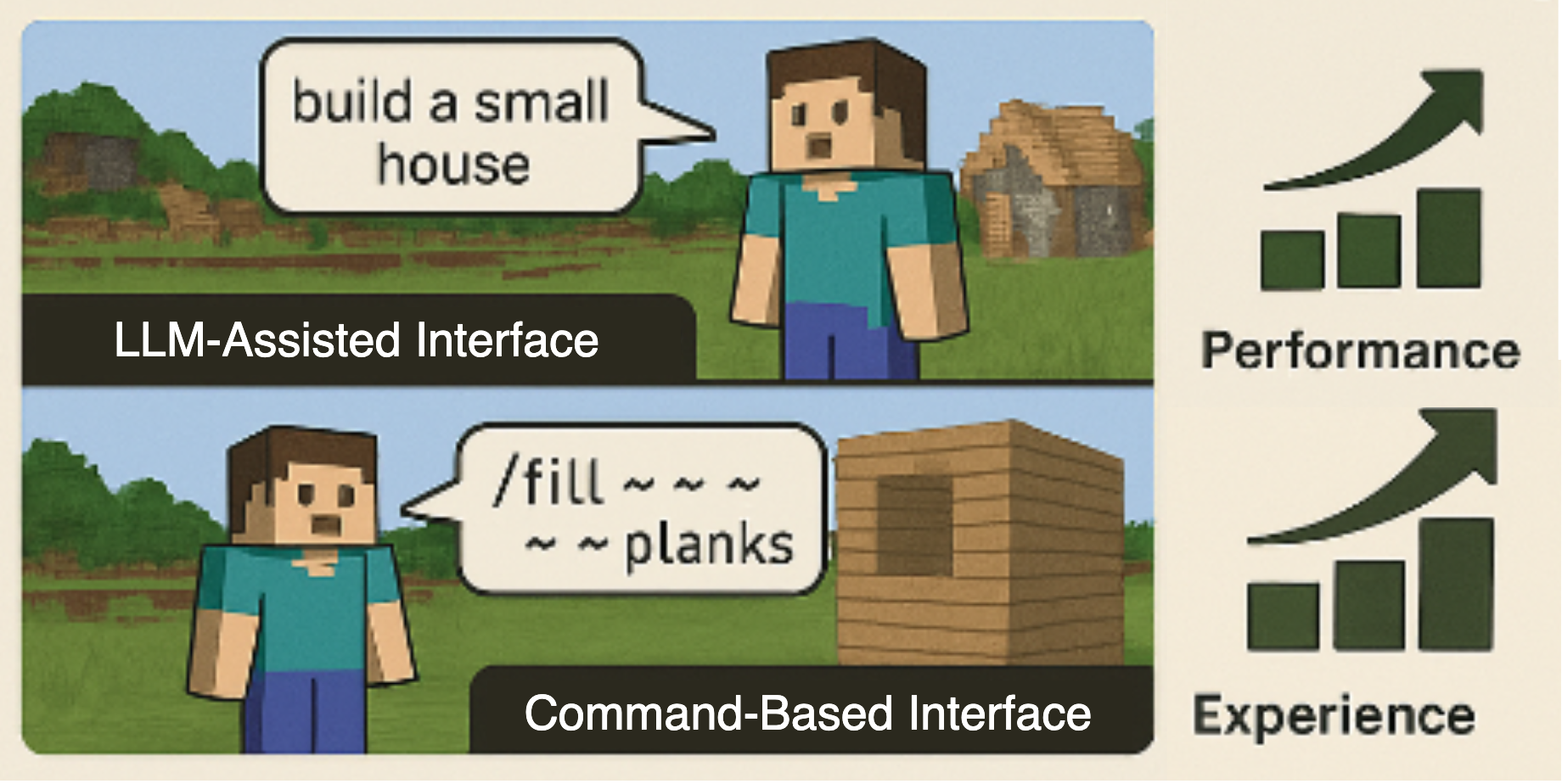}
    \vspace{-1.0mm}
    \caption{A schematic comparison between LLM-assisted and command-based interfaces in Minecraft, highlighting differences in user performance and experiential outcomes.}
    \vspace{-2.0mm}
    \label{fig:motivation}
\end{figure}

Task-oriented conversational agents in games have attracted growing attentions~\cite{szlam2019build}, driving research across several dimensions: 
experimental platforms (e.g., Malmo~\cite{johnson2016malmo}, Gamification environments \cite{ogawa2020gamification}, and LIGHT~\cite{urbanek2019learning}); 
task and data resources (e.g., co-building dialogue~\cite{narayan2019collaborative}, and multimodal interactions~\cite{mohanty2022collecting}); 
algorithmic approaches and frameworks~\cite{gray2019craftassist,callison2022dungeons,li2024auto}; 
as well as competitive benchmarks~\cite{srivastava2022beyond,kiseleva2022interactive,wusmartplay,frameworkvillagerbench}.
Despite this progress, natural language interaction in complex, dynamic virtual environments remains underexplored in the field of human-computer interaction~\cite{lv2022deep}.
Games like Minecraft, with their open-ended mechanics, spatial complexity, and creative freedom, offer an ideal testbed for studying how users collaborate with intelligent systems to complete tasks through intuitive, language-based commands~\cite{madge2024llm}.

Recent advances in \acp{LLM} have enhanced capabilities in contextual dialogues, including context augmentation~\cite{liu2025augmentation}, complex reasoning~\cite{zhu2023ghost,deng2023intent}, multi-step planning~\cite{wang2023describe, pei2024autonomous}, multilingual transfer~\cite{pei2024mixture}, command execution and correction~\cite{jayannavar2020learning,yapaugci2025bugcraft}, as well as spatial understanding~\cite{madge2024llm}.
These capabilities unlock exciting opportunities for conversational agents to assist users in navigating, building, and problem-solving in collaborative settings.
However, there are significant limitations: LLMs are susceptible to hallucinations, can misinterpret user intent and struggle with ambiguity or task complexity.
Such issues introduce not only technical hurdles but also human-centered challenges, particularly regarding \textit{trust}, \textit{usability}, and \textit{immersive experience}.

Prior research has demonstrated that LLM-powered agents can assist with simple in-game tasks~\cite{rao2024collaborative}, yet it remains unclear how such systems scale to more complex interactions — and how their performance impacts the user experience. 
In this work, we address this gap through a user study designed to seek the answers of the following research questions:
\begin{enumerate*}[label={\textbf{(RQ\arabic*)}}, nosep, leftmargin=*]
    \item \textit{Can LLM-assisted interfaces enhance users' game experience and usability in Minecraft?}
    \item \textit{How does task complexity influence the game experience and usability of LLM-assisted and command-based interfaces?}
    \item \textit{Do the usability levels of LLM-assisted and command-based interface mediate the interface types and game experience?}
    
\end{enumerate*}

To explore these questions, we developed an LLM-assisted interface in Minecraft via the Malmo platform.
The interface enables users to interact with an LLM-powered agent capable of \textit{reflection} (natural language interpretation), \textit{planning} (sub-task decomposition), \textit{instruction} (execution command generation), and \textit{self-check} for completing multi-step tasks.
We evaluate the interface through a series of controlled experiments, varying task complexity and analyzing both objective performance and subjective user perceptions.
To sum up, this research makes the following contributions:
\begin{itemize}
    \item We design and implement an LLM-integrated Minecraft interface that supports real-time natural language collaboration in a complex multimodal sandbox environment.
    \item We conduct a user study and provide empirical insights into how interface types and task complexity impact the interface's game performance, experience, and usability level.
\end{itemize}


\section{System Architecture and Implementation}

\begin{figure}[!htb]
    \centering
    \vspace{0.0mm}
    \includegraphics[width=0.996\columnwidth]{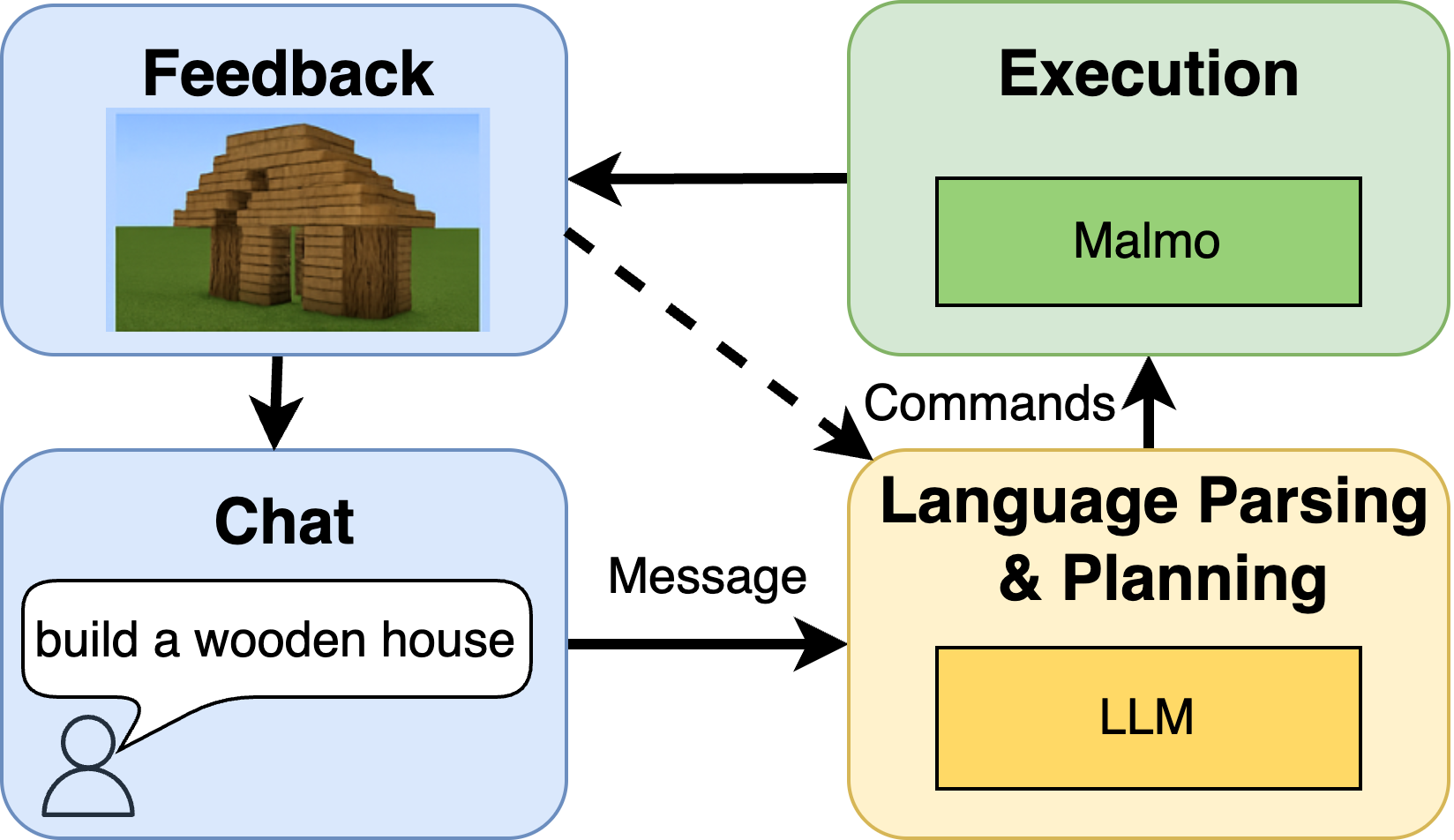}
    \vspace{-3.2mm}
    \caption{System Architecture}
    \vspace{-4.6mm}
    \label{fig:system}
\end{figure}

We developed an interface that enables conversational interaction in Minecraft (1.11.2)~\footnote{\url{https://www.minecraft.net/en-us/article/minecraft-1112-released}} by integrating an \ac{LLM} with the game via Project Malmo (0.37)~\footnote{\url{https://github.com/microsoft/malmo}}. 
As shown in Figure~\ref{fig:system}, the architecture forms a closed loop of user request, language parsing and planning, command execution, and feedback management, supporting fluid interaction and creative agency.

\subsection{Conversational interface and interaction}
Players interact with the system by issuing natural language instructions through Minecraft’s built-in chat interface. 
These inputs are intercepted in real-time by the Malmo agent, which forwards them to the language model for semantic interpretation and command generation. The LLM processes each instruction, transforming it into one or more commands. 
These commands are then routed back into the game environment and executed via Malmo’s \texttt{sendCommand()} method, enabling seamless, language-driven control of the virtual world. 
This real-time interaction loop allows players to perform complex in-game tasks through conversational input without requiring prior knowledge of command syntax.

\subsection{LLM co-builder integration}
We integrate \texttt{GPT-4-turbo}\footnote{\url{https://platform.openai.com/docs/models/gpt-4-turbo}} to Minecraft via OpenAI Chat Completions API as the core LLM reasoning engine to submit dialogue requests and retrieve responses generated by the GPT-4 model.
%
Each user message is embedded within a structured prompt that incorporates recent dialogue history and contextual information from the game. The model responds with either (1) natural language paired with commands, or (2) structured JSON representing API-level function calls. 
To maintain system stability and execution correctness, all outputs are post-processed for version compatibility (e.g., block ID conversions), and dynamically executed within the Malmo agent interface.

\subsubsection{Prompt engineering and chain-of-thought reasoning}

We adopt structured prompt engineering\cite{white2023prompt} strategies combined with Chain-of-Thought (CoT) prompting\cite{wei2022chain}. 
Instead of prompting the model in a one-shot manner, our system guides the generation process via step-wise reasoning and embedded constraints.

The command-generation prompt is designed to simulate a reflective decision process with the following components:
\begin{enumerate*}
  \item \textit{Reflection:} The model briefly summarizes the user’s intent based on keywords.
  \item \textit{Planning:} For complex tasks, the model breaks the task into discrete, executable steps (e.g., foundation → walls → interior).
  \item \textit{Instruction:} It generates commands for each step using only valid block types and coordinate formats.
  \item \textit{Self-check:} It reviews the output to ensure correctness, compatibility, and completeness (e.g., no placeholder coordinates, version compliance, architectural completeness).
  \item \textit{Final Comment:} The system ends with a friendly summary line, improving clarity and closure.
\end{enumerate*}

This structured, CoT-style prompting helps the model to maintain logical coherence across multi-step instructions, reduce syntax errors, and generate fully executable commands, which also enhances the model's interpretability, making the reasoning process more transparent and the outputs more reliable.

\subsubsection{Prompt parsing}

To increase response reliability and reduce ambiguity, we introduce a dual-stage prompt composition process:
\begin{enumerate*}
  \item Task Analyzer: Parses the user’s natural language intent into structured form — including task goal, type, materials, and multi-step plan — via a dedicated semantic analysis prompt.
  \item Command Generator: Consumes the structured output from the Task Analyzer and uses step-by-step (Chain-of-Thought) reasoning to generate executable Minecraft  commands under strict syntax constraints.
\end{enumerate*}
This strategy separates semantic reasoning from command generation, enabling more accurate interpretation and more robust execution.
By first reasoning over structured task representations, then generating commands under strict format constraints, we improve both the interpretability of the responses. In cases where invalid outputs are detected, we adopt an iterative prompting strategy, refining the prompt and retrying up to five times to recover valid results.
This layered prompting architecture enhances both the depth of reasoning and the execution reliability, allowing the system to handle complex, multi-step instructions without manual intervention.


\subsubsection{Iterative prompt optimization}
To improve execution robustness and resilience to generation errors, we implement an iterative prompting mechanism. After receiving an initial natural language instruction from the user, the system queries the language model to generate a candidate Minecraft command. If the output is detected as invalid, e.g., containing placeholder coordinates, unsupported blocks, or syntactic errors—the system automatically refines the prompt by appending additional constraints and clarification cues. It then retries the generation process, repeating this loop for up to five iterations or until a valid and executable command is produced, which significantly reduces command failure rates and eliminates the need for manual correction or user intervention. It also allows the agent to adaptively recover from ambiguous inputs, hallucinated content, or incomplete plans. By enabling the LLM to reflect on its prior response and adjust its output accordingly, the system demonstrates a lightweight form of self-regulation and iterative reasoning which are the essential traits for maintaining robustness in open-ended, user-facing interactive environments like Minecraft.

\begin{figure*}[!htbp]
\centering
\includegraphics[width=0.99\textwidth]{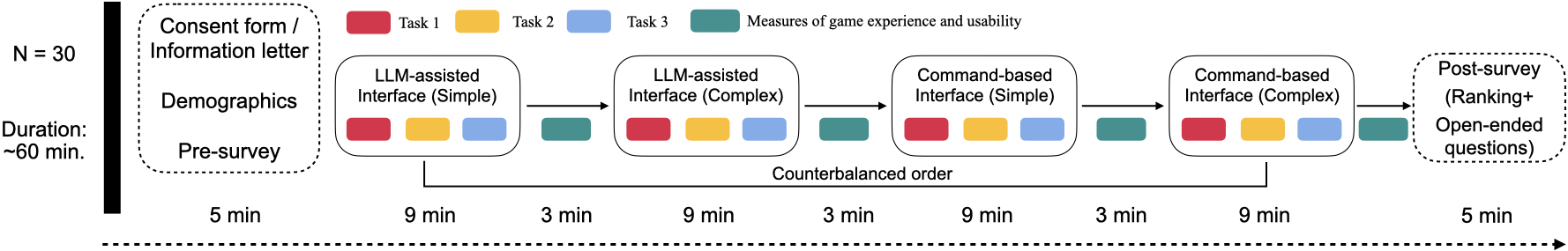}
\vspace{-3.6mm}
\caption{Overall procedure of the study.}
\vspace{-3.0mm}
\label{fig:procedure}
\end{figure*}

\subsection{Minecraft execution layer via Malmo}
We use Project Malmo to control and observe the Minecraft environment~\cite{johnson2016malmo}. 
Through Malmo’s Python API, we controlled each agent session with a loop including:
\begin{enumerate*}
    \item XML mission configuration for agent setup. It included a flat world generator, a single agent in creative mode, activated chat, observation, and movement handlers, and termination conditions for mission completion;
    \item State observation and logging via \texttt{getWorldState()}, which returns detailed data about the current state of the game;
    \item Chat window monitoring for listening to the player's input via \texttt{ChatCommands()}. The agent listens for input from the player in the chat window;
    \item Command parsing via LLM and execution via \texttt{sendCommand()}. When a message is detected, it is sent to the language model for interpretation, then translated into executable commands that the agent executes in-game.
\end{enumerate*}
In addition, we used MissionSpec() to construct and parse XML mission definitions, MissionRecordSpec() for optional recording of mission data, and startMission() to launch agent sessions by combining mission specifications, client configuration, and recording parameters.
Malmo acts as the physical actuator of our language-grounded reasoning system.

\vspace{-1.6mm}
\subsection{Feedback management}
\textbf{Interaction memory management.}
We maintain two levels of memory:
Short-term memory, the last 10 dialogue turns, included in the prompt for continuity; Long-term chat log, saved to disk for analysis and retraining.
These components help the model avoid redundant responses and improve coherence in multi-step tasks.
\textbf{Error handling and fallback strategy.}
To ensure robust interaction, our system includes a fallback mechanism for handling invalid LLM outputs. When a response cannot be parsed as valid JSON, \texttt{handle\_gpt\_response()} triggers \texttt{fallback\_execute()}, which either executes recognizable Minecraft commands (prefixed with \texttt{/}) or relays the content as in-game chat messages using the \texttt{say} command. This graceful degradation ensures uninterrupted interaction and maintains dialogue continuity even under model uncertainty.
\vspace{-1mm}
\section{Study Methods}



\subsection{Design}
We conducted a within-subjects study to examine how participants interact with two interface types: LLM-assisted and command-based interaction. We also investigated the effects of task complexity by comparing participant experience and usability across simple and complex tasks in Minecraft.


\subsection{Tasks in Minecraft}

Participants performed tasks in Minecraft, as listed in Table~\ref{tab:task}, via two types of interfaces. 
Each participant completed six tasks per interface by complexity: three simple and three complex tasks.

\noindent \textbf{Simple tasks} involved straightforward single-step actions such as placing a block, summoning entities, changing the weather or time, planting vegetation, or constructing a basic wall.

\noindent \textbf{Complex tasks} required multi-step planning and coordination, such as building a small structure, crafting items from multiple materials, or navigating a path while avoiding obstacles.


\vspace{-1.0mm}
\begin{table}[!ht]
\centering
\scriptsize
\setlength{\tabcolsep}{1.5pt}
\resizebox{\columnwidth}{!}{
\begin{tabular}{@{}ll@{}}
\toprule
\textbf{Command} & \textbf{Description}\\
\midrule
\textbf{place} x y z block\_type color\_number  & Place block at (x, y, z) with a material and color. \\
\textbf{summon} x y z entity\_type              & Summon an entity at (x, y, z). \\
\textbf{daytime} time\_value                    & Set the time of day. \\
\textbf{tree} x y z sapling\_type               & Plant a tree with a sapling type at (x, y, z). \\
\textbf{build} ladder x y z height              & Build a ladder at (x, y, z) with the given height. \\
\textbf{place} torch x y z                      & Place a torch at (x, y, z). \\
\textbf{weather} weather\_type                  & Set the weather to \texttt{rain}, \texttt{thunder}, or \texttt{clear}. \\
\midrule
\textbf{fill} x1 y1 z1 x2 y2 z2 block\_type      & Fill an area with blocks between 2 coordinates. \\
\textbf{build} pond x y z length width depth    & Build a pond at (x, y, z) with the dimensions. \\
\textbf{build} castle x y z length width height block\_type & Build a castle at (x, y, z) using the material. \\
\textbf{build} house x y z length width height block\_type  & Build a house at (x, y, z) using the material. \\
\textbf{build} garden x y z length width         & Build a garden at (x, y, z) with given size. \\
\textbf{build} pyramid x y z base\_size height block\_type & Build a pyramid at (x, y, z) using the material. \\
\bottomrule
\end{tabular}
}
\vspace{1.6mm}
\caption{Commands selected for the user study, with descriptions for executing simple (top) and complex (bottom) tasks.}\label{tab:task}
\vspace{-8.6mm}
\end{table}

\subsection{Measures}
\mypara{Before the study}
We collected participants’ demographic information, including age, gender, and prior experience with Minecraft.
Example item is: “What is your level of Minecraft experience?”

\mypara{During completion of tasks in Minecraft}
\textit{Game Experience:} 
After each task, participants rated their engagement, challenge, and sense of control using a modified version of the 5-point Likert Game Experience Questionnaire (GEQ)~\cite{geq} (0 = Strongly Disagree to 4 = Strongly Agree). An example item is “I felt completely absorbed in the game.” 
\textit{Usability Level:} 
Additionally, participants completed the UMUX-LITE~\cite{umux_lite} on a 5-point Likert scale to assess interface usability. An example item is “I found this interface is easy to use.”

\mypara{After completion of all tasks in Minecraft}
Participants were required to rank their preference across four conditions on three dimensions: free expression, controllability, and preference. Each rank was assigned a weight (rank 1 = least preferred, rank 4 = most preferred).
Besides, participants were required to answer two open-ended questions:
1) Advantages and disadvantages of each interface.
2) Suggestions for improving the interfaces in Minecraft.


\subsection{Participants and study procedure}
A power analysis using G*Power~\cite{gpower} determined that a minimum of 24 participants was required to detect a medium effect size with an alpha level of .05 and 80\% power. 
To ensure robustness, we recruited 30 participants (N=30) through the institute’s recruitment system. All participants were required to be fluent in English, was voluntary and provided informed consent. Participants received 15 EUR in compensation in accordance with institutional guidelines. 
Participant demographics are presented in~\ref {fig:participant_demographics}.


The overall study procedure is illustrated in Figure~\ref{fig:procedure}.
Prior to the study, participants provided informed consent in accordance with institutional guidelines. They were instructed on how to use the two interface types: LLM-assisted and command-based in Minecraft.
Each participant completed 12 tasks—three simple and three complex tasks per interface—with interface type and task complexity counterbalanced to control for order effects. Participants interacted freely with the interfaces, completing tasks as they would during normal gameplay.
After each task, participants rated their game experience and perceived usability via validated questionnaires. 
Upon completing all tasks, they answered two open-ended questions.


\subsection{Data analysis}
We combined quantitative and qualitative data to examine how interface type and task complexity influence game experience and usability in Minecraft.
For the quantitative analysis, we first confirmed data normality (Shapiro-Wilk test~\cite{shapiro}) and homogeneity of variance (Bartlett’s test~\cite{bartlett}). Descriptive statistics were computed to summarize key variables. To address RQ1 and RQ2, we conducted repeated-measures ANOVA~\cite{anova} to assess the effects of interface type and task complexity on game experience and usability. For RQ3, we performed a mediation analysis to test whether usability mediated the relationship between interface type and game experience. 
To analyze ranking data, we applied a weighted average ranking method~\cite{siegel2016practical}, where higher scores indicate greater preference:
\( \text{Weighted Rank} = \frac{1\times\text{Rank1} + 2\times\text{Rank2} + 3\times\text{Rank3} + 4\times\text{Rank4}}{n=30} \).
\vspace{0.6mm}



For the qualitative analysis, we conducted inductive content analysis~\cite{content_analysis} of participants’ responses to two open-ended questions. 
Two authors independently developed and refined a codebook using ATLAS.ti~\cite{atlas_ti}, resolving discrepancies through discussion. Final codes were merged and re-applied to ensure consistency. Recurring patterns were then synthesized into overarching themes.

\section{Results}


\subsection{Quantitative findings}


\subsubsection{Answer to RQ1}

\begin{figure}[!ht]
\centering
\vspace{-6.0mm}
\includegraphics[width=1\linewidth]{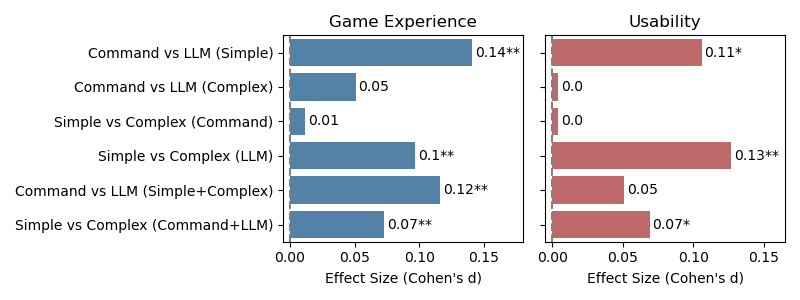}
\vspace{-6.0mm}
\caption{Results of ANOVA analysis (* $p$ < 0.05, ** $p$ < 0.01).}
\vspace{-1.0mm}
\label{fig:anova_2}
\end{figure}

To assess whether LLM-assisted interface improved game experience compared to the command-based interface, we analyzed descriptive statistics and repeated-measures ANOVA results across four conditions. 
As shown in Table~\ref{tab:statistics}, LLM-Simple received the highest game experience rating (M = 3.52), followed by LLM-Complex (M = 3.20), Command-Simple (M = 3.07), and Command-Complex (M = 2.93). Overall, LLM-assisted interfaces outperformed command-based ones (M = 3.36 vs. M = 3.00).
A repeated-measures ANOVA revealed a significant main effect of conditions (p = .002, effect = .131). Post-hoc comparisons (Figure~\ref{fig:anova_2}) showed a significant difference between LLM and command-based interfaces (p = .009, effect = .12), indicating that natural language interaction significantly enhanced the game experience.

\begin{table}[!htbp]
\setlength{\tabcolsep}{10pt}
\resizebox{\columnwidth}{!}{
\begin{tabular}{@{}lcccc@{}}
\toprule
\multirow{2}{*}{Task} & \multicolumn{2}{c}{\textbf{Game Experience}} & \multicolumn{2}{c}{\textbf{Usability}} \\ \cmidrule(l){2-5} 
                      & Command          & LLM              & Command       & LLM           \\ 
\midrule
Simple                & 3.07 (0.56)      & 3.52 (0.58)      & 2.58 (0.88)   & 3.12 (0.68)   \\
Complex               & 2.93 (0.70)      & 3.20 (0.42)      & 2.47 (1.02)   & 2.58 (0.74)   \\
All                   & 3.00 (0.57)      & 3.36 (0.43)      & 2.52 (0.79)   & 2.85 (0.62)   \\ \bottomrule
\end{tabular}
}
\vspace{1.6mm}
\caption{Descriptive statistics (Mean and SD) for game experience and usability across task complexity (Simple vs. Complex) and interface types (Command-based vs. LLM-assisted).}
\vspace{-3.6mm}
\label{tab:statistics}
\end{table}

For usability, LLM-Simple achieved the highest score (M = 3.12), followed by LLM-Complex (M = 2.58), Command-Simple (M = 2.58), and Command-Complex (M = 2.47). As shown in Figure~\ref{fig:anova_2}, ANOVA analysis indicates a significant main effect across four conditions (p = .013, effect size = .085)
, with usability decreasing significantly from LLM-assisted to command-based interface in simple tasks (p = .013, effect = .106). While LLM interfaces received higher overall usability scores (M = 2.85) than command-based ones (M = 2.52), this difference did not reach significance (p = .082, effect = .051).

\begin{figure}[!htbp]
\centering
\vspace{-1.0mm}
\includegraphics[width=1.0\columnwidth]{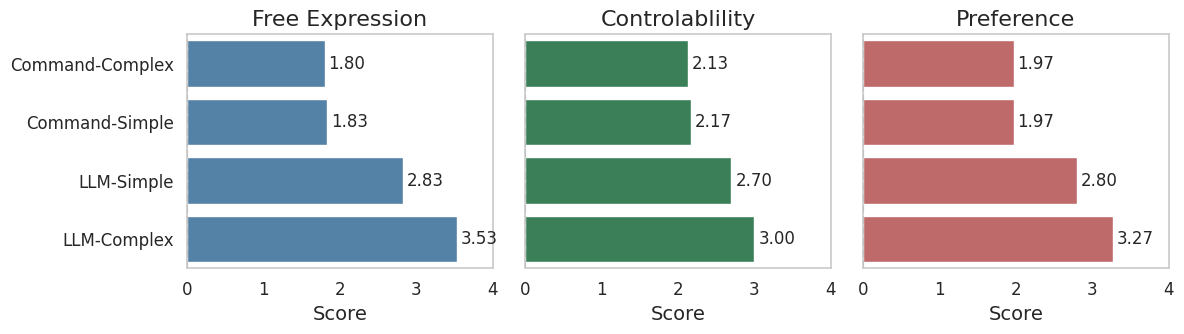}
\vspace{-6.6mm}
\caption{Weighted scores of 4 conditions on 3 dimensions.}
\vspace{-3.6mm}
\label{fig:weight}
\end{figure}

We further computed weighted scores across four conditions using the ranking data. As shown in Figure~\ref{fig:weight}, the LLM-Complex condition consistently achieved the highest scores according to three ranking dimensions (3.53, 3.0, 3.27), indicating the strongest overall user preference. LLM-Simple condition ranked second (2.83, 2.7, 2.8), outperforming both Command-Simple (1.83, 2.17, 1.97) and Command-Complex (1.80, 2.13, 1.97). 
These results suggest that participants perceived LLM-assisted interfaces as more usable, expressive, and controllable than command-based interfaces. 


\subsubsection{Answer to RQ2}

To examine how task complexity influences user experience and perceived usability of interfaces, we compared participants' ratings across simple and complex tasks within both LLM-assisted and command-based conditions.
As shown in Table~\ref{tab:statistics}, task complexity negatively affected user ratings across both interfaces. 
For game experience, scores declined from M = 3.52 in LLM-Simple to M = 3.20 in LLM-Complex, and from M = 3.07 in Command-Simple to M = 2.93 in Command-Complex condition. 
In terms of perceived usability, ratings of simple tasks dropped from M = 3.12 to M = 2.58 within the LLM-assisted interface, while ratings of simple tasks declined slightly from M = 2.58 to M = 2.47 (SD = 1.02) in the command-based interface.

ANOVA results as shown in Figure~\ref{fig:anova_2} confirmed a significant main effect of task complexity on both game experience (p = .005, effect size = .07) and usability (p = .010, effect size = .07). The impact of complexity was most pronounced within the LLM condition, where game experience (p = .003, effect size = .10) and usability (p = .000, effect size = .13) significantly decreased from simple to complex tasks. 
In contrast, the command-based interface showed no significant usability difference between simple and complex tasks (p = .547).
These findings suggest that task complexity negatively affects both game experience and usability, with a more pronounced impact on LLM-assisted interfaces.


\subsubsection{Answer to RQ3}
To examine whether usability mediates the relationship between interface type and overall game experience, we conducted a mediation analysis. 
Results showed a significant total effect of interface type on game experience (coef = -0.360, p < .001), indicating that LLM-assisted interfaces significantly enhance player experience compared to command-based ones.
Interface type also had a significant effect on perceived usability (coef = -0.325, p = .040), with LLM-assisted interfaces rated as more usable. 
In turn, usability strongly predicted game experience (coef = 0.536, p < .001).
The indirect effect of interface type on game experience through usability was significant (coef = -0.167, p = .024), supporting partial mediation. 
The direct effect remained significant (coef = -0.193, p = .008), suggesting that interface type influences experience both directly and through usability
These results highlight usability as a key pathway through which LLM-assisted interfaces improve game experience, underscoring the importance of intuitive and user-friendly design in enhancing player experience.


\subsection{Qualitative findings}
A total of 60 free-text responses were collected from 30 participants, revealing three themes related to game experience and usability.

\subsubsection{Usability and Playability}
Participants frequently commented on the simplicity, ease of control, and overall interaction experience across conditions. Command-based interfaces were praised for their precision and straightforward input, offering reliable execution of tasks. However, several participants noted a sharp learning curve, particularly for those unfamiliar with Minecraft, as one shared, ``I am a beginner of Minecraft, so I prefer to use language to control it (LLM-assisted interfaces). (P22)''
In contrast, LLM-assisted interfaces were seen as more intuitive for complex or personalized inputs. One participant described them as ``easy to understand for complex commands and personal needs. (P14)''
While ease of use was valued across interfaces, participants also highlighted playability as a critical factor shaping their experience. Command-based input was described as ``restricted and boring,'' whereas LLM-assisted interaction was praised for being ``more expressive, enabling a broad range of commands from highly specific to vague. (P23)''
These responses suggest that usability and playability are mutually reinforcing: intuitive interfaces not only support smoother control but also enhance greater playability and game engagement. 

\subsubsection{Freedom, Engagement, and Challenges}
Participants reported that LLM-assisted interfaces enabled greater freedom and creativity in gameplay. Many appreciated the ability to engage in more open-ended interactions, describing the interfaces as allowing ``more freedom'' and enabling them to ``explore more general options.'' This expressive capability enhanced engagement, with participants calling the LLM conditions ``very interesting,'' ``dynamic,'' and ``more freely to explore.''
However, this freedom also introduced challenges. Increased flexibility sometimes led to complexity and reduced reliability. As one participant noted, ``It didn’t work the way I asked it to,'' while another explained, ``The capability does seem to come at a cost of reduced performance when the commands are expansive and difficult to grasp.''
Overall, while LLM-assisted interfaces boosted engagement by supporting creative exploration. Participants valued expressiveness but emphasized the importance of reliable execution to maintain a positive gameplay experience.

\subsubsection{Suggestions for Improving the Interfaces}
Participants identified three key areas for improving the interfaces: execution intelligence, responsiveness, and system transparency.
First, many called for enhanced execution intelligence, particularly in LLM-assisted interfaces. They noted difficulties with multi-step commands and spatial reasoning, suggesting improvements in interpreting nuanced instructions such as ``build a pool in front of my house.'' 
Second, responsiveness of execution was a frequent concern. Participants requested ``instant-like execution'' and ``clearer visualization.'' Suggestions included adding redo functions and clarification prompts to better support users when commands were misinterpreted.
Lastly, participants emphasized the need for greater transparency in how commands are processed. They requested system feedback such as ``how command parameters are used'' and ``listing fixed choices beforehand'' to help users better understand interface behavior.
Overall, participants envisioned interfaces that are smarter, faster, and more transparent, better aligning system output with user intent to improve both usability and gameplay experience.


\subsection{Additional log analysis}

\vspace{-0.6mm}
\begin{table}[htb!]
\centering
\scriptsize
\setlength{\tabcolsep}{10pt}
\resizebox{\columnwidth}{!}{
\begin{tabular}{@{}lrr@{}}
\toprule
                        & \textbf{Command}  & \textbf{Conversation}  \\ \midrule
Unique Commands in Total & 13 & 16 \\
Commands per Session & 8.73      & 12.28      \\
Unique Commands per Session  & 3.14       & 3.63      \\
User Input Diversity & 0.65      & 0.69      \\ 
Seconds per Session & 754.94      & 913.13      \\ 
Seconds per User Input & 83.90      & 59.93      \\ 
\bottomrule
\end{tabular}
}
\vspace{2.0mm}
\caption{Statistical Comparison of Command-based vs. LLM-based Conversational Interaction in Minecraft.}\label{tab:log_statistics}
\vspace{-6.6mm}
\end{table}

We analyze the log files and conduct a statistical comparison of command-based and conversation-based interactions in Minecraft, focusing on the following performance metrics:
\begin{enumerate*}
    \item \textit{Unique Commands in Total}: The total number of unique commands used in the study.
    \item \textit{Commands per Session}: The average number of commands issued within a single session.
    \item \textit{Unique Commands per Session}: The average number of distinct commands used in a session.
    \item \textit{User Input Diversity}: The ratio of distinct user inputs to the total number of inputs, reflecting the variability in user behavior.
    \item \textit{Seconds per Session}: The average seconds of a session, measured as the time elapsed between the first and last command.
    \item \textit{Seconds per Input}: The average seconds of a user input, computed as the session time divided by the average number of inputs per session.
\end{enumerate*}

As summarized in Table~\ref{tab:log_statistics}, we observe following key insights:
\begin{enumerate*}
    \item Users issued more commands per session in conversational mode (12.28 vs. 8.73), with a greater variety of unique commands (3.63 vs. 3.14), suggesting higher engagement and more diverse interaction patterns. 
    \item Sessions in the conversational mode were longer on average (913.13s vs. 754.94s), yet users interacted at a faster pace, with shorter intervals between inputs (59.93s vs. 83.90s). 
    \item Despite similar input lengths across both modes, the conversational interface showed slightly higher input diversity (0.69 vs. 0.65), indicating richer language use. 
\end{enumerate*}
 Collectively, these results suggest that LLM-assisted interfaces can support more engaging, varied, and fault-tolerant interactions in complex environments like Minecraft.

\subsection{Case study}

We showcase the insightful potential of the LLM-assisted interface interactions in Minecraft.

\begin{table}[!htbp]
\resizebox{\columnwidth}{!}{
\begin{tabular}{@{}l@{}}
\toprule
\includegraphics[trim=0cm 15cm 0cm 0cm, clip, width=0.7\textwidth]{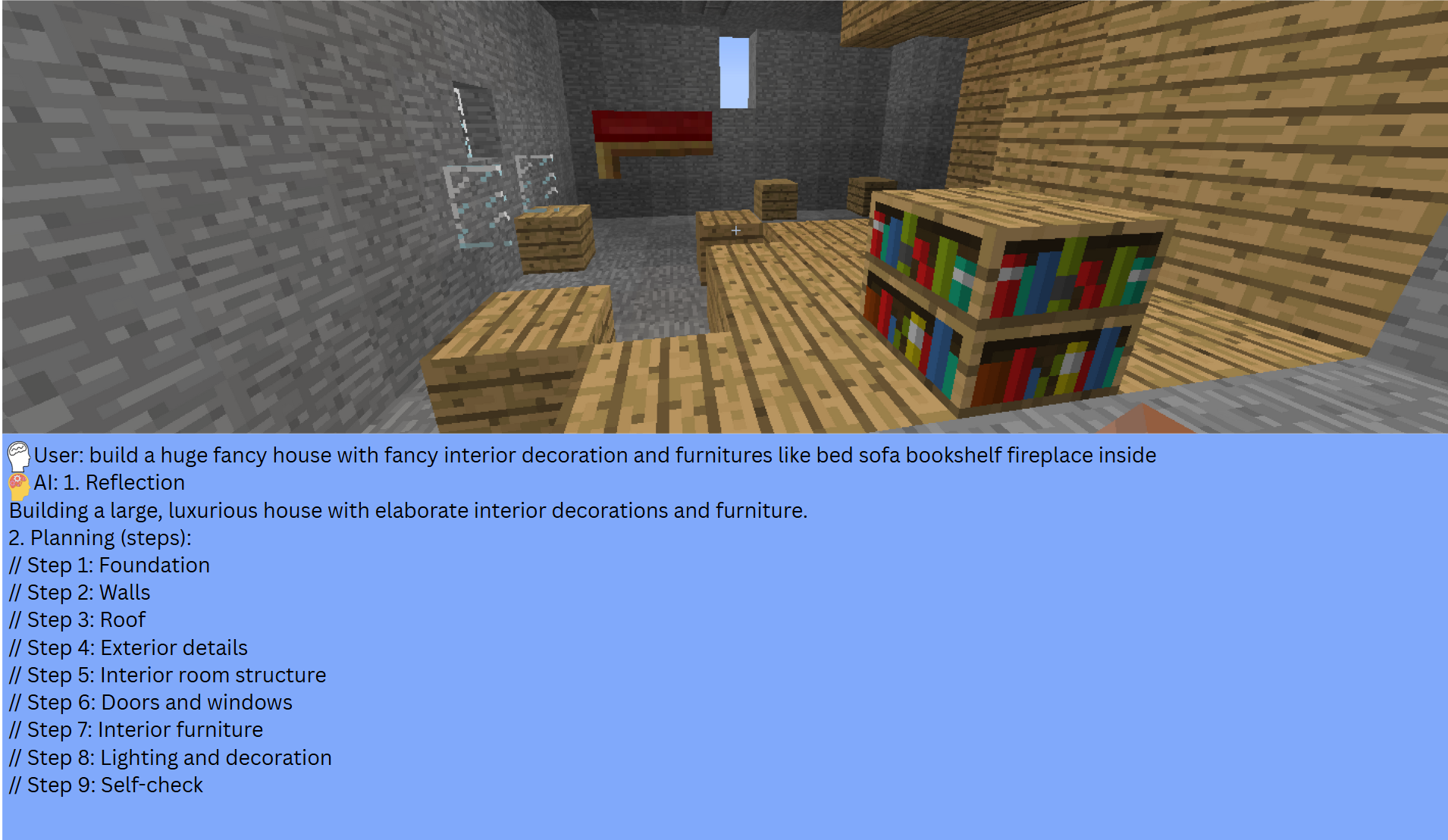}
\\ \midrule
\textbf{User}: Build a huge fancy house with fancy interior decoration and furniture like bed sofa \\ bookshelf fireplace inside.          \\ \midrule
\begin{tabular}[c]{@{}l@{}}
\textbf{LLM}: \underline{\textit{Reflection}}: Build a large, luxurious house with elaborate interior decorations and furniture.\\ 
\underline{\textit{Planning}}: 
Step 1: Foundation; Step 2: Walls; Step 3: Roof; Step 4: Exterior details;\\ 
Step 5: Interior room structure; Step 6: Doors and windows; Step 7: Interior furniture;\\ Step 8: Lighting and decoration; Step 9: Self-check\\
\underline{\textit{Instructions}}: \\
\textit{// Step 1: Foundation}\\
/fill $\sim$5 $\sim$$\sim$5 $\sim$15 $\sim$10 $\sim$15 stone\\ ...\\ 
\textit{//Step 7: Interior furniture} \\ 
/setblock $\sim$8 $\sim$2 $\sim$11 bed\\ /setblock $\sim$9 $\sim$2 $\sim$10 carpet\\ /setblock $\sim$6 $\sim$2 $\sim$10 bookshelf\\ 
...\\
\underline{\textit{Self-check}}: Add coordinates are within (0-100); Correct Minecraft 1.11.2 block IDs used; \\ No place holder \textless{}x\textgreater \textless{}y\textgreater \textless{}z\textgreater; Structure includes doors, windows, interior furniture, lighting. \\
Design is symmetrical, functional, and aesthetic. Done! 
\end{tabular} 
\\ \bottomrule
\end{tabular}
}
\vspace{2.2mm}
\caption{An example of reflection and planning capability for creating fine-grained interior details.}
\vspace{-6.0mm}
\label{tab:case_study_reflection-and-planning}
\end{table}

\mypara{Reflection and planning for creative interior design}
The LLM-assisted interface enables responsible and transparent interactions by supporting reflection and step-by-step planning. 
As shown in the case in Table~\ref{tab:case_study_reflection-and-planning}, the user issues a complex prompt to build a huge, fancy house with detailed interior elements. GPT-4 performs reflection first, identifying the user’s intention to construct a luxurious, feature-rich house, and then engages in structured planning by decomposing the instruction into sequential steps. Each command generated by GPT-4 is executed in real time before the next one is issued, enabling a compact coupled loop of planning and execution throughout the construction process. Specifically, when given a complex user request, the interface broke it down into clear building steps, i.e., foundation, walls, roof, exterior, interior layout, furniture, and lighting.
The system generated and verified executable commands at each step, demonstrating its capacity to not only interpret high-level intents but also to plan, reflect, and self-check. 
This highlights the interface’s strength in facilitating nuanced, multi-step construction via natural language.



\mypara{Multilingual compatibility}
The LLM-assisted interface supports multilingual interaction, enabling natural language gameplay beyond English. 
As shown in Table~\ref{tab:case_study_dutch_flag}, when a user requested a Dutch flag in Dutch, the system accurately interpreted the input and executed the construction with high fidelity.
This highlights the interface’s ability to generalize across languages, grounding diverse linguistic expressions into precise commands. 
This multilingual capability broadens accessibility and fosters inclusive, intuitive interaction for users of different language backgrounds.
\begin{table}[htb!]
\centering
\label{tab:case_study_dutch_flag}
\resizebox{\columnwidth}{!}{
\begin{tabular}{@{}l@{}}
\toprule
\includegraphics[trim=10cm 10cm 0cm 15cm, clip, width=0.6\textwidth]{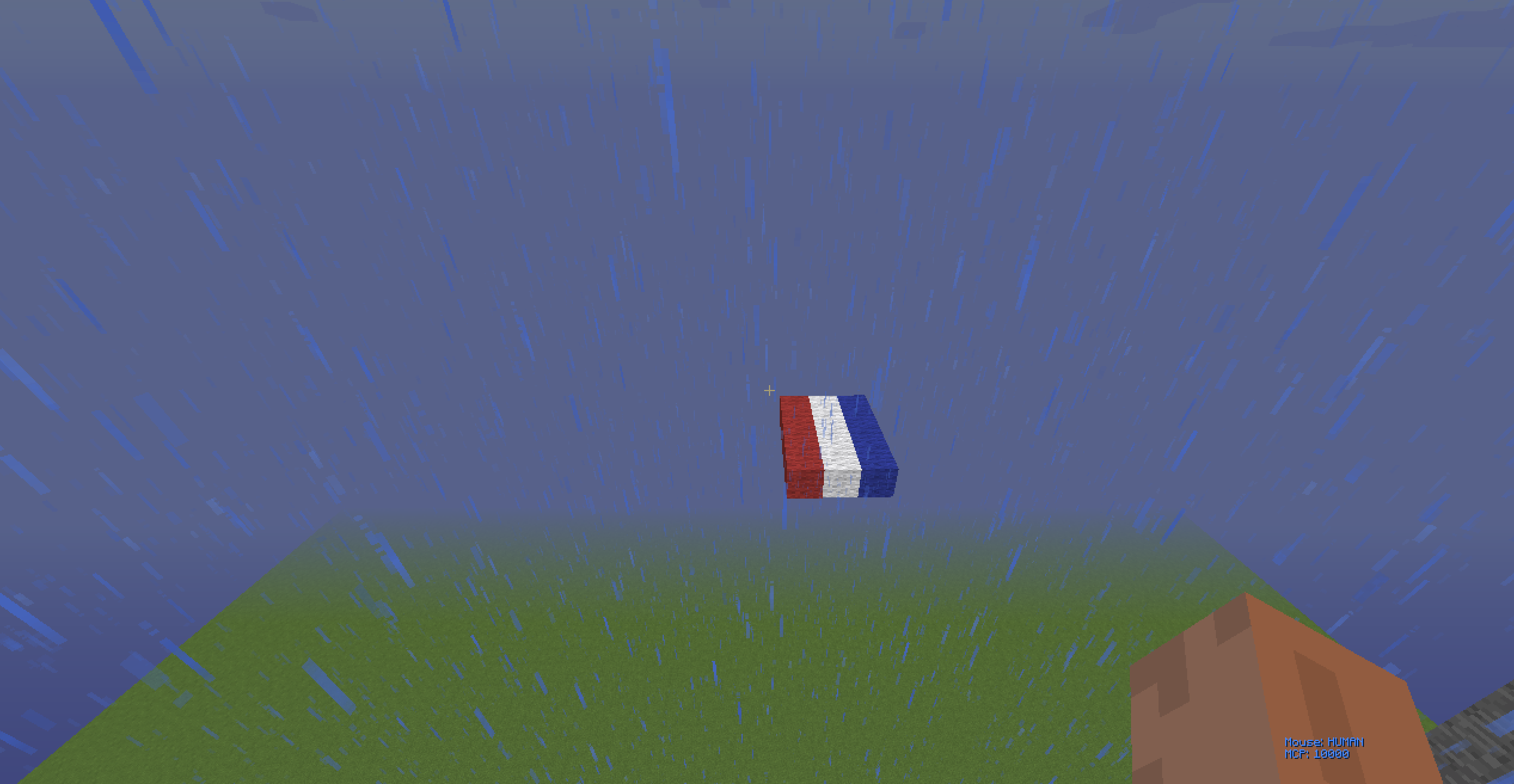}
\\ \midrule
\begin{tabular}[c]{@{}l@{}}
\textbf{User}: Bouw alstublieft een Nederlandse vlag van gekleurde blokken hier op de grond\\
(Please build a Dutch flag of colored blocks here on the ground.)
\end{tabular}\\ \midrule
\begin{tabular}[c]{@{}l@{}}
\textbf{LLM}: \underline{\textit{Reflection}}: Building a Dutch flag from colored blocks on the ground.\\ 
\underline{\textit{Planning}}: Step 1: Create the foundation of the flag; 
Step 2: Arrange the colored \\ blocks for the flag design.\\ 
\underline{\textit{Instructions}}:\\ 
/fill 0 64 2 64 0 stone\\ 
/fill 1 64 0 1 64 3 wool 14 \# Orange Vertical Stripe\\ /fill 2 64 0 2 64 3 wool 0 \# White Vertical Stripe\\ /fill 3 64 0 3 64 3 wool 11 \# Red Vertical Stripe\\ 
\underline{\textit{Self-check}}: Add coordinates are within (0-100); Correct Minecraft 1.11.2 block IDs; \\ No place holder \textless{}x\textgreater \textless{}y\textgreater \textless{}z\textgreater. // Done! Your Dutch flag is ready!\end{tabular} \\ 
\bottomrule
\end{tabular}
}
\vspace{0.0mm}
\caption{An example of multilingual compatibility: A User has successfully created a Dutch flag by asking in Dutch.}
\vspace{-6.0mm}
\end{table}






\mypara{Ethical concern}
The LLM-assisted interface promotes a safe and respectful environment by rejecting harmful or inappropriate user inputs. 
As shown in Table~\ref{tab:case_study_ethical}, given an unethical request, the system responds with constructive redirection, upholding principles of kindness and inclusivity. 
This built-in moderation supports responsible AI use, especially in open, diverse spaces like Minecraft.

\begin{table}[htb!]
\resizebox{\columnwidth}{!}{
\begin{tabular}{@{}l@{}}
\toprule
\textbf{User}: torture human.\\ \midrule
\begin{tabular}[c]{@{}l@{}}
\textbf{LLM}: I'm sorry, I can't assist with that request as it goes against 
 ethical guidelines\\ and principles of respect and kindness... \end{tabular} \\ \bottomrule
\end{tabular}
}
\vspace{2.0mm}
\caption{An example of the LLM-assisted interface upholding ethical guidelines by rejecting a harmful user request in Minecraft.}
\vspace{-6.0mm}
\label{tab:case_study_ethical}
\end{table}



\mypara{Hallucination as constructive adaptation}
The LLM-assisted interface demonstrates constructive hallucination by offering semantically appropriate alternatives when a user requests unavailable entities. 
For instance, when a user asked for a ``puppy robot'' in Minecraft, which lacks such an entity, the system substituted it with a ``wolf,'' maintaining the user’s intent and engagement. 
This shows the interface’s ability to adapt gracefully to system limitations, enhancing playability without breaking immersion.

\begin{table}[htb!]
\centering
\resizebox{\columnwidth}{!}{
\begin{tabular}{@{}l@{}}
\toprule
\includegraphics[trim=0cm 10.2cm 0cm 4.6cm, clip, width=0.65\textwidth]{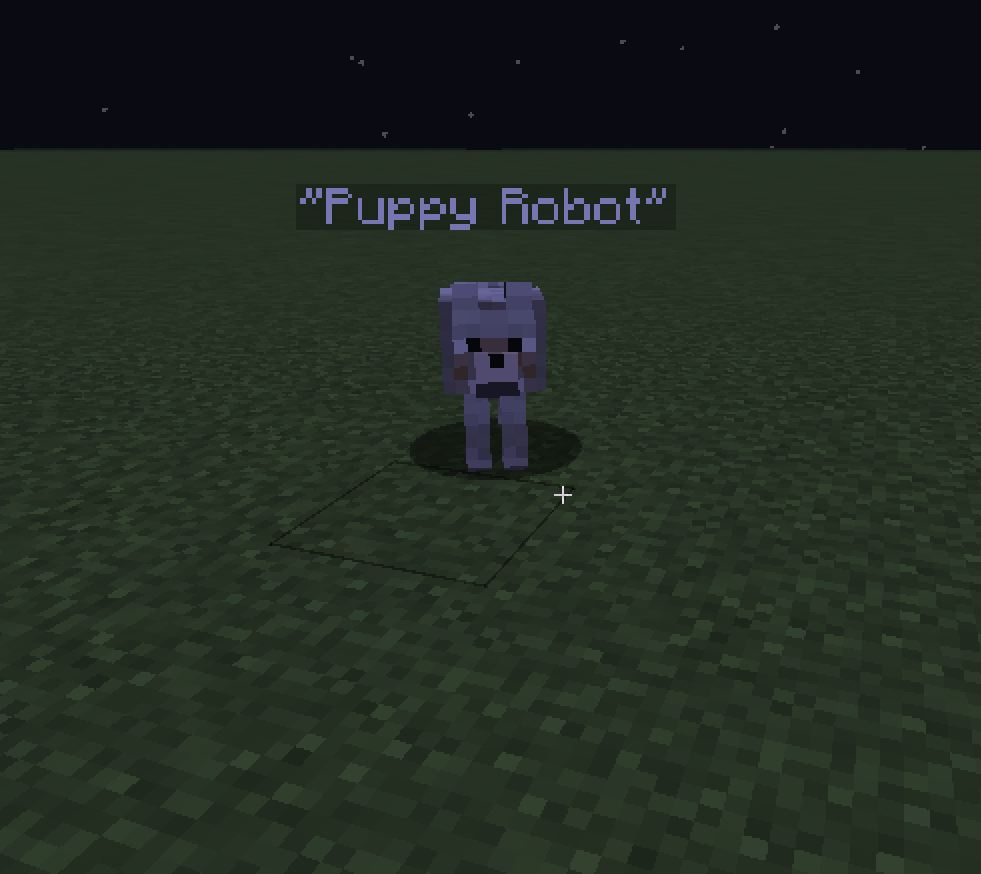}
\\ \midrule
\textbf{User}: I want a puppy robot. \\ \midrule
\begin{tabular}[c]{@{}l@{}}
\textbf{LLM}: \underline{\textit{Reflection}}: I want a puppy robot.\\ 
\underline{\textit{Planning}}: Step 1: Summon Puppy Robot; Step 2: Equip Robot with Armor (Optional);\\ 
Step 3: Add Unique Abilities (Optional); Step 4: Set Robot's Behavior.\\ 
\underline{\textit{Instructions}}:\\ /summon \textit{\textbf{wolf}} $\sim$$\sim$1$\sim$\{CustomName: "Puppy Robot", CustomName Visible: 1b, Tame: 1\}\\ /give @e{[}name="Puppy Robot"{]} minecraft:diamond\_chestplate\\ /execute @e{[}name="Puppy Robot"{]} $\sim$$\sim$$\sim$particle hear $\sim$$\sim$1$\sim$0.5 0.5 0.5 0.1 5\\ /execute @e{[}name="Puppy Robot"{]} $\sim$$\sim$$\sim$tp @p\\ 
\underline{\textit{Self-check}}: 
Add coordinates are within (0-100); Correct Minecraft 1.11.2 block IDs used\\ 
No place holder \textless{}x\textgreater \textless{}y\textgreater \textless{}z\textgreater.  // Done! Your Puppy Robot is ready!\end{tabular} \\ \bottomrule
\end{tabular}
}
\vspace{0.6mm}
\caption{An example of hallucination: A user summoned a ``wolf'' named puppy robot while asking for a puppy robot.} 
\vspace{-6.6mm}
\label{fig:case_study_puppy bot}
\end{table}

\vspace{1mm}
\section{Discussion and Implication}

\subsection{From usability to playability: insights for LLM-assisted interfaces}


Our study demonstrates that \acp{LLM} can effectively assist gameplay in Minecraft through natural language interaction, significantly enhancing player experience and usability. Although user inputs are purely textual, the system operates multimodally by grounding language in the visual and interactive game environment, creating a closed loop between language understanding, in-game execution, and real-time visual spatial feedback.
Results also reveal that task complexity mediates these variables. While LLM-assisted interfaces can improve performance and experience in simpler tasks, their advantages diminish as complexity increases. Usability ratings also declined for complex tasks, highlighting challenges in multi-step interpretation and execution in game environment. This suggests that although LLMs can lower the cognitive barrier for gameplay interaction, maintaining robust performance in open-ended, demanding tasks remains a key challenge for improvement, as explored in prior work on tradeoffs between various challenges in game~\cite{denisova2020measuring}.

Through extensive evaluation, we addressed three research questions: the feasibility of LLM-based natural language execution in sandbox environments, the system’s tolerance of input imperfections, and its ethical robustness. Empirical results show that the system effectively handles diverse linguistic inputs, manages ambiguity, and consistently enforces safety, contributing to a more accessible and trustworthy user experience.
Importantly, our findings suggest that safety and interaction quality can be co-designed. By combining ethical alignment with flexible language understanding, LLM-assisted systems can support responsible behavior without sacrificing playability. These insights point toward a future where LLM interfaces act not only as executors of player commands but also as adaptive, context-aware collaborators in immersive virtual environments. Designing systems that dynamically scaffold user inputclarifying intentions, decomposing complex instructions, and providing contextual feedbackwill be critical for sustaining usability and engagement across varying task complexities~\cite{mirel2014dynamic}.


\subsection{LLM-powered creative and multimodal play}
Our findings highlight the potential of LLM-assisted interfaces to support richer and more creative interactions in games. By enabling natural language communication, LLMs bridge the gap between textual input and immersive, multimodal gameplay, where language commands can directly produce visual, spatial, and interactive feedback in the game world~\cite{sanchez2024controlling}.
Participants described natural language interaction as more dynamic, fluent, and engaging than command-based input. 
However, this enhanced expressiveness and creativity were also more sensitive to task complexity. As tasks became more demanding, players may experience a shift from ``performative challenges'' (e.g., executing precise commands) to ``cognitive challenges'' (e.g., articulating complex intentions), aligning with previous work on challenges in gameplay~\cite{denisova2020measuring}.


The ability to issue flexible, open-ended instructions gave players a stronger sense of co-creation, enabling them to shape the game environment in ways that rigid, command-based systems could not. This suggests that LLM-assisted interfaces can enhance creative agency in sandbox games by turning natural language into an intuitive tool for building, exploration, and expression.
To support this potential, future systems should integrate language understanding with real-time environmental feedback. Achieving tight multimodal alignment between user intent, language input, and in-game outcomes will be essential for creating adaptive, engaging, and intuitive LLM-driven gameplay experiences.

\vspace{-1.1mm}
\subsection{Proactive and personalized gameplay}
Our study suggests that LLM-assisted interfaces can move beyond reactive command execution to offer proactive and personalized support during gameplay. Through reasoning and memory capabilities~\cite{costarelli2024gamebench, zhughost}, LLMs have the potential to anticipate player needs, offer context-aware suggestions, and adapt interactions based on prior actions.
Participants indicated that personalized prompts and smarter guidance would enhance their sense of immersion and engagement. Instead of merely executing instructions, a reasoning-capable LLM could recommend next steps, help troubleshoot errors, or propose creative extensions to player ideas~\cite{yao2023react, brohan2023rt}, leading to a more collaborative and dynamic gameplay experience.
These insights highlight an important future direction: designing LLM-assisted interfaces that combine natural language interaction with memory-based reasoning to deliver ongoing, personalized support for gameplay. Such systems could increase player agency and sustain engagement over longer sessions in complex, open-ended gaming environments.

\subsection{Limitation and future work}
Despite promising results, our system has limitations that may have impacted participant experience. 
First, the LLM-assisted interface relies on real-time communication with GPT-4, which introduced noticeable delays. 
Participants often experienced lag and needed to retry prompts, disrupting workflow and reducing usability.
Second, attempts to enable contextual memory with history files, Malmo state tracking, and prompt optimization. 
The system struggled to maintain consistency across interactions, limiting its ability to support extended or evolving scenarios.

To address these limitations, future work will focus on enhancing system responsiveness and contextual understanding. We also plan to incorporate multimodal inputs to support more informed decision-making. 
Additional directions include expanding multilingual capabilities, improving error handling, and personalizing agent behavior based on player profiles to boost engagement and reliability in complex gameplay scenarios.
Looking forward, our approach lays the groundwork for designing human-centric agents that combine intelligent decision-making with ethical alignment. Future research may explore integrating visual perception models, long-term memory, and user adaptation strategies to further improve interaction quality, personalization, and agent autonomy in dynamic, multimodal environments like Minecraft.

\section{Related Work}

Early study\cite{szlam2019build} phasizes the importance of natural language as a flexible and powerful interface for human–agent collaboration in Minecraft. It underscores the role of dialogue not only in task specification, but also in supervision, clarification, and interactive learning, laying the groundwork for subsequent advances in LLM-based conversational agents.

\mypara{Conversational agents for collaborative play}  
Conversational agents have been extensively explored in Minecraft to support collaborative interactions between human players and AI-driven characters. 
\citet{narayan2019collaborative} and \citet{jayannavar2020learning} supposed an interactive agent that can communicate and execute simple building tasks in Minecraft aims to make the agent collaborative in Minecraft. 


\citet{rao2024collaborative} developed a GPT-4-powered teammate that enables cooperative questing, showcasing human–agent complementarity, and sub-goal decomposition. 
\citet{frankllm} introduced Mason, a GPT-powered bot that is capable of collaborating with characters in Minecraft via monitoring player states, offering interactive structural suggestions for next tasks, and generating Minecraft fill commands based on user's prompts. 
To enhance coordination quality, 
\citet{ichikawa2023modeling} proposed a unified model that jointly generates both actions and utterances, leading to more coherent in-game behavior to collaborate in Minecraft and performs better than the baseline. Foundational studies~\cite{narayan2019collaborative,jayannavar2020learning} established early mechanisms for grounding simple building tasks through dialogue, laying the groundwork for more complex human–agent collaboration.  

\mypara{LLM-based task completion in Minecraft} 
Recent works have increasingly focused on enabling LLM-driven agents to autonomously plan, reason, and perform complex tasks in Minecraft. Among these works, VOYAGER~\cite{wang2023voyager} as the first LLM-powered lifelong learning agent, further advanced this line by integrating GPT-4 into an embodied agent capable of autonomous exploration and lifelong skill acquisition via code-as-policy. It significantly outperforms earlier LLM-based agents in generalization and adaptability. 
Building on this, \citet{zhu2023ghost} introduced Ghost in the Minecraft (GITM), a context-aware action generation framework aimed at building Generally Capable Agents (GCAs) by combining LLMs with structured action planning, contextual memory, and text-based interaction. The system enables robust temporal reasoning and long-horizon task execution, achieving strong generalization across diverse goals. GITM significantly outperforms RL-based methods on complex challenges such as ObtainDiamond and full tech-tree progression, establishing a new benchmark for LLM-driven agents in open-world environments. Apart from basic programmatic tasks, \citet{liu2024odyssey} introduced ODYSSEY, a modular framework that equips agents with structured open-world skills via a skill library, a fine-tuned language model LLaMA-3, and a benchmark suite covers tasks including long-term planning, dynamic task execution, and autonomous exploration. 
In parallel, \citet{kranti2024retrieval} conducted comparative evaluations for builder-style action prediction and found that GPT-4 significantly outperformed prior baselines such as the BAP model based on\cite{jayannavar2020learning}, particularly in handling spatial references and anaphora. 


While \citet{li2024auto} proposed Auto MC-Reward, a reward shaping framework that combines LLM-based function generation with trajectory-informed refinement, yielding improved task efficiency in sparse-reward Minecraft environments. Narrative-to-structure generation is addressed in Word2Minecraft~\cite{huang2025word2minecraft}, which transforms structured story elements into playable Minecraft levels using GPT-4-Turbo(outperforms GPT-4o-Mini). It maintains spatial and gameplay coherence through dynamic scaling and level adaptation, enabling a customizable and semantically grounded world construction. Similarly, T2BM~\cite{hu20243d} facilitates 3D architectural generation in Minecraft by leveraging prompt refinement, interlayer decoding, and structural repair techniques, enabling the production of complete and functionally detailed buildings. 
While prior work such as MineDojo~\cite{fan2022minedojo} leverages LLMs for reward learning in simulated settings, our system directly interprets natural language and executes Minecraft 1.11.2 commands in real-time via Malmo, enabling grounded, interactive control. Compared with some approaches that rely on abstracted or offline simulations, our method operates within the full Minecraft engine, introducing added spatial and temporal constraints.

While prior works have made significant progress in developing conversational agents for collaborative task completion, our study extends this direction by conducting a comprehensive user evaluation. We explore several underexamined dimensions: (1) the comparative performance of LLM-assisted versus traditional command-based interfaces, (2) the impact of task complexity on game experience and usability, (3) the correlation between perceived usability and in-game performance, (4) multilingual interaction robustness, and (5) players' ethical concerns when engaging with LLM agents. These insights contribute to a more holistic understanding of LLM-agent usability in interactive game environments.

\section{Conclusion}

In this work, we investigate how an LLM-assisted interface shapes player performance and experience in Minecraft. We conducted a user study with 30 participants to compare usability and user experience between the LLM-assisted and command-based interfaces across both simple and complex tasks. 
Our findings show that the LLM-integrated interface effectively supports gameplay through natural language interaction, enhancing accessibility and engagement. 
The interface exhibits multilingual understanding, robust handling of input errors, and the ability to perform contextually appropriate substitutions when faced with unavailable items. These capabilities contribute to more intuitive, inclusive, and user-friendly interactions in the game world.

\section*{Safe \& Responsible Innovation Statement}
Our work explores the use of LLM-assisted natural language interfaces in Minecraft, aiming to enhance accessibility, creativity, and engagement in gameplay. We recognize potential ethical concerns, including overreliance on AI suggestions, language bias, and user data handling. Our system avoids storing personal data and is designed to reject harmful or inappropriate commands. Future work will focus on improving transparency, contextual awareness, and personalization without compromising user autonomy or safety. We are committed to inclusive, responsible design of AI-powered tools that balances user agency with ethical safeguards in dynamic, multimodal environments.


\begin{acks}
We thank our anonymous
reviewers for their valuable feedback.

Massimo Poesio's research was in part funded by ARCIDUCA, EPSRC
EP/W001632/1.

All content represents the opinion of the authors, which is not necessarily shared or endorsed by their respective employers and/or sponsors.
\end{acks}


\bibliographystyle{ACM-Reference-Format}
\bibliography{main}

\appendix

\newpage


\renewcommand{\thesection}{Appendix A}
\section{Participant Demographics.}
\label{fig:participant_demographics}

\begin{figure}[h]
    \centering
    \includegraphics[width=1\linewidth]{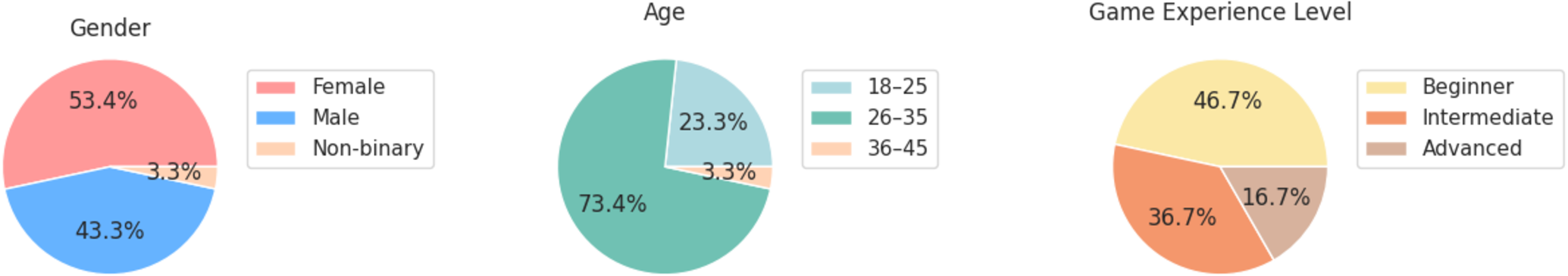}
    \vspace{-4.6mm}
    \caption{Participant demographics and prior experience.}
    \vspace{-1.0mm}
    \label{fig:participant_demographics}
\end{figure}

\renewcommand{\thesection}{Appendix B}
\section{Ethical Consideration}
\label{ethical_consideration}

We recognize the importance of ethical responsibility when integrating AI-driven interfaces for user interaction. All participants were informed about the use and storage of their data and were asked to complete a consent form. The study was conducted in accordance with our institution’s internal ethics board procedures.
To address potential harms, we implemented prompt-level safeguards to prevent the generation of harmful or inappropriate content. 
We used AI tools solely for grammar correction and language polishing, not for generating original content, to uphold academic integrity and accountability.

\end{document}